\documentclass[10pt,english]{article}
\usepackage[T1]{fontenc}
\usepackage[latin9]{inputenc}
\usepackage{geometry}
\usepackage{color}
\usepackage{float}
\usepackage{amsmath}
\usepackage{graphicx}
\usepackage{setspace}

\makeatletter
\newcommand{\lyxaddress}[1]{
\par {\raggedright #1
\vspace{1.4em}
\noindent\par}
}

\usepackage{cite,times}
\date{}
\renewcommand{\textendash}{--}

\makeatother

\usepackage{babel}
\begin{document}

\title{Entanglement concentration protocols for GHZ-type entangled coherent
state based on linear optics}

\author{Mitali Sisodia$^{1}$\thanks{mitalisisodiyadc@gmail.com; }, Chitra
Shukla$^{2}$\thanks{Corresponding Author: shuklac@pcl.ac.cn; chitrashukla07@gmail.com}}
\maketitle

\lyxaddress{\begin{center}
$^{1}$ Indian Institute of Technology Jodhpur, Jodhpur, 342037 India
\\
$^{2}$Center for Quantum Computing, Peng Cheng Laboratory, Shenzhen
518055, People\textquoteright s Republic of China
\par\end{center}}
\begin{abstract}
We proposed two entanglement concentration protocols (ECPs) to obtain
maximally entangled Greenberger-Horne-Zeilinger (GHZ)-type entangled
coherent state (ECS) from the corresponding partially entangled GHZ-type
ECSs. We obtained the first ECP using a partially entangled GHZ-type
ECS assisted with a superposition of single-mode coherent state, however
the second ECP is designed using two copies of partially entangled
GHZ-type ECSs. The success probabilities have also been calculated
and discussed for both the ECPs. We have further compared the success
probabilities of our first ECP for 3-mode GHZ-type ECS with an ECP
of 3-mode W-type ECS and found that our ECP is more efficient (maximal
success probabilities) for larger value $(\beta=0.7)$ of state parameter.\textcolor{red}{{}
}For the physical realization, two optical circuits (for two ECPs)
using linear optical elements, viz 50:50 beam splitter, phase shifter,
and photon detectors are provided, which support the future experimental
implementation possible with the present technology. 
\end{abstract}

\section{Introduction\label{sec:Introduction}}

The robust quantum entanglement among the distant communicating parties
is the backbone for any realization of the quantum information tasks
like quantum teleportation \cite{tele}, dense coding \cite{dense_coding},
quantum key distribution \cite{ekert}, quantum key agreement \cite{QKA1,QKA2},
quantum secret sharing \cite{QSS}, hierarchical quantum communication
\cite{HQIS,HJRSP,Ba_An_Hierarchical_QT} and quantum secure direct
communication \cite{review,QSDC-Review}. The sharing of maximal entanglement
plays a key role in the success of the above applications to be performed
deterministically \cite{Pathak_Book}. In reality, the entanglement
is generated locally and then distributed (shared) among the communicating
users located remotely. Unfortunately, while transmission, manipulation
and storage process, the entangled quantum systems undergo the interaction
with an unavoidable noisy environment and hereof experience the degradation
in maximal entanglement which lead to the probabilistic communication
\cite{Probabilistic_Teleporation}. Therefore, a maximally entangled
state (MES) will be turned into a less (partial) entangled state or
even a mixed state thereby reduced (less than unity) teleportation
fidelity would be obtained, which not only puts the security of quantum
communication/computation \cite{Nielsen_Chuang} tasks at a high risk
but also limits the distance of entanglement distribution. To overcome
such serious problems, two prevalent approaches have been proposed
to convert a non-MES into a MES in order to achieve the unit teleportation
fidelity for the successful performance of the quantum information
processing tasks. They are named as entanglement concentration protocol
(ECP) and entanglement purification protocol (EPP). The ECP is considered
for achieving the desired MES out of an ensemble of pure non-MESs
whereas EPP is to obtain a high quality entangled states from the
corresponding mixed states. In 1996, the first ECP has been proposed
by Bennett et al. based on Schmidt decomposition \cite{ECP_Bennet}.
Afterwards, several ECPs have been reported for different quantum
states \cite{ECP_Bose,ECP_Sheng2,Zhao_Cluster_ECP,ECP_Zhao,Our_ECP1,Long_ECP_cluster,ECP_Sheng_2019,Wang_2019,ECP_Bell_like_N_GHZ_N_W,ECP_Experiment}
{[}and references therein{]}. 

A Greenberger-Horne-Zeilinger (GHZ) state \cite{GHZ-W-Class} is one
of the important quantum entangled state among them. Along with the
state generation \cite{Generation_GHZ_W,Generating_GHZ}, GHZ state
has been popular for the realization of various applications \cite{Experimental_GHZ-6-photon_nonlocality,GHZ-Paradox,Quan_2-copy_teleportation}
in discreet variable (DV) domain where the entanglement is encoded
in polarization of photons. For the perfect execution of such GHZ
state applications numerous ECPs have also been introduced recently
\cite{ECP_Bell_like_N_GHZ_N_W,ECP_GHZstate,ECP_GHZ_type_photons,ECP_Bell_GHZ}
{[}and references therein{]}. Interestingly, in the similar way, GHZ
state also has its significant importance in continuous variable (CV)
domain where the entanglement is encoded in superposition of coherent
states $|\pm\alpha\rangle$ ($\alpha$ being amplitude with large
values). Specifically, much attention has been given to continuous
variable states in theory \cite{Barry2,Barry_2} and experimental
preparations \cite{Experimental_CV_Cluster1,Experimental_CV_Cluster}
due to their quantum nonlocality and high capacity. Practically, the
CV states are superior as they overcome the limitations govern by
DV states, such as low detection efficiency, high cost, and some limitations
related to preparation and manipulation.

Therefore, beyond the conventional approach of using DV, the CV states
have been of high interest among the researchers in recent years.
Entangled coherent state (ECS) is a type of CV state that has attracted
a lot of attention \cite{Barry2,Barry_2}. Further, among the generation
of various ECSs (in theory and experiment) like W-type ECS \cite{Ba_AN_generation_GHZ-type_ECS},
Cluster-type ECS \cite{Experimental_CV_Cluster1,Experimental_CV_Cluster,Ba_An_teleportation_CTECS},
the generation schemes of GHZ-type ECS have also been put forward
in theory and experiment \cite{Experimental_CV_Cluster1,Ba_AN_generation_GHZ-type_ECS}.
Moreover, many applications in quantum communication \cite{Ba_An_teleportation_CTECS,Chen__teleportation_CTECS}
using Cluster-type ECS, quantum cryptography using two-mode squeezed
states \cite{CV_QKD}, quantum teleportation using 3-mode ECS \cite{QT_with_3mode_ECS_PRA,Improving_QT_with_3mode_ECS_PRA}
and quantum key distribution using GHZ-type ECS \cite{QKD_GHZ_ECS},
quantum computation \cite{QC_CV_Cluster1,QC_CV_Cluster2} using Cluster-type
ECS and quantum information processing tasks \cite{CVQKD2} using
W-type ECS \cite{Nguan_Ba_An} have been proposed so far, that have
reflected the significance of ECSs \cite{Experimental_CV_Cluster1,Experimental_CV_Cluster,Ba_AN_generation_GHZ-type_ECS,Ba_An_teleportation_CTECS}
and motivated the recently reported CV-based ECPs \cite{ECP_Bell_type_ECS,W-ECS,Our_OECP}
(and references therein), too. In particular, Sheng et al. proposed
ECP for Bell-type ECS \cite{ECP_Bell_type_ECS} and W-type ECS \cite{W-ECS}.
Subsequently, Mitali et al. \cite{Our_OECP} designed two ECPs for
Cluster-type ECS based on linear optics. Independently, the hybrid
ECPs have also attracted the recent interest of the researchers \cite{HES_Cluster_ECP,Our_Hybrid_ECP_2020}.
However, yet no ECP has been designed specifically for GHZ-type ECS.
In fact, several applications have been proposed exploiting the GHZ-type
ECS such as quantum teleportation \cite{QT_with_3mode_ECS_PRA,Improving_QT_with_3mode_ECS_PRA},
quantum key distribution \cite{QKD_GHZ_ECS}, very recently two communication
schemes for controlled versions of quantum teleportation \cite{H_Prakash_2019_CQT}
and entanglement diversion \cite{H_Prakash_2019_CED}. It would be
apt to note that with the recent progress of applications in quantum
communication and cryptography \cite{QT_with_3mode_ECS_PRA,Improving_QT_with_3mode_ECS_PRA,QKD_GHZ_ECS,H_Prakash_2019_CQT,H_Prakash_2019_CED},
it is natural that the ECP has been in demand for GHZ-type ECS. Hence,
designing ECPs for this state would be worth and of practical interest
for the experimental implementation of these applications \cite{H_Prakash_2019_CQT,H_Prakash_2019_CED}
and as well as for more potential applications based on GHZ-type ECS.
With this motivation, here we aim to propose two ECPs for 3-mode GHZ-type
ECS, and to the best of our knowledge these ECPs have been the first
attempt for 3-mode GHZ-type ECS.

The rest of the paper is organized as follows. In Sec. \ref{sec:ECP-for-GHZ-type},
we have described our proposed ECP for 3-mode GHZ-type ECS with the
help of ancillary single-mode coherent state. Further, in Sec. \ref{sec:ECP-with-two}
our next ECP has been discussed which uses two copies of 3-mode GHZ-type
ECSs. Subsequently in Sec. \ref{sec:Success-probability}, the success
probabilities have been calculated and their corresponding plot has
been shown in Fig. \ref{fig:(Color-online)-Variation}. Finally, the
work has been concluded in the end in Sec. \ref{sec:Conclusion}.

\section{\textcolor{black}{ECP for} partially entangled 3-mode \textcolor{black}{GHZ-type
ECS with the help of a superposition of single-mode coherent state\label{sec:ECP-for-GHZ-type}}}

\textcolor{black}{A maximally entangled }3-mode G\textcolor{black}{HZ-type
ECS can be expressed as }

\textcolor{black}{
\begin{equation}
\begin{array}{ccc}
|\Psi\rangle_{abc} & = & N_{0}\left[|\alpha\rangle_{a}|\alpha\rangle_{b}|\alpha\rangle_{c}+|-\alpha\rangle_{a}|-\alpha\rangle_{b}|-\alpha\rangle_{c}\right],\end{array}\label{eq:maximally-GHZ_ECS}
\end{equation}
where $N_{0}=\left[2\left(1+2e^{-6|\alpha|^{2}}\right)\right]^{-\frac{1}{2}}$
is the normalization coefficient. }$\left|\pm\alpha\right\rangle $
are the coherent states and $\alpha$ is the amplitude with large
values. The subscripts $a$,\textcolor{black}{{} $b$ and $c$ represents
the states possessed by Alice, Bob and Charlie, respectively}\textcolor{blue}{. }

\begin{figure}
\begin{centering}
\includegraphics[scale=0.7]{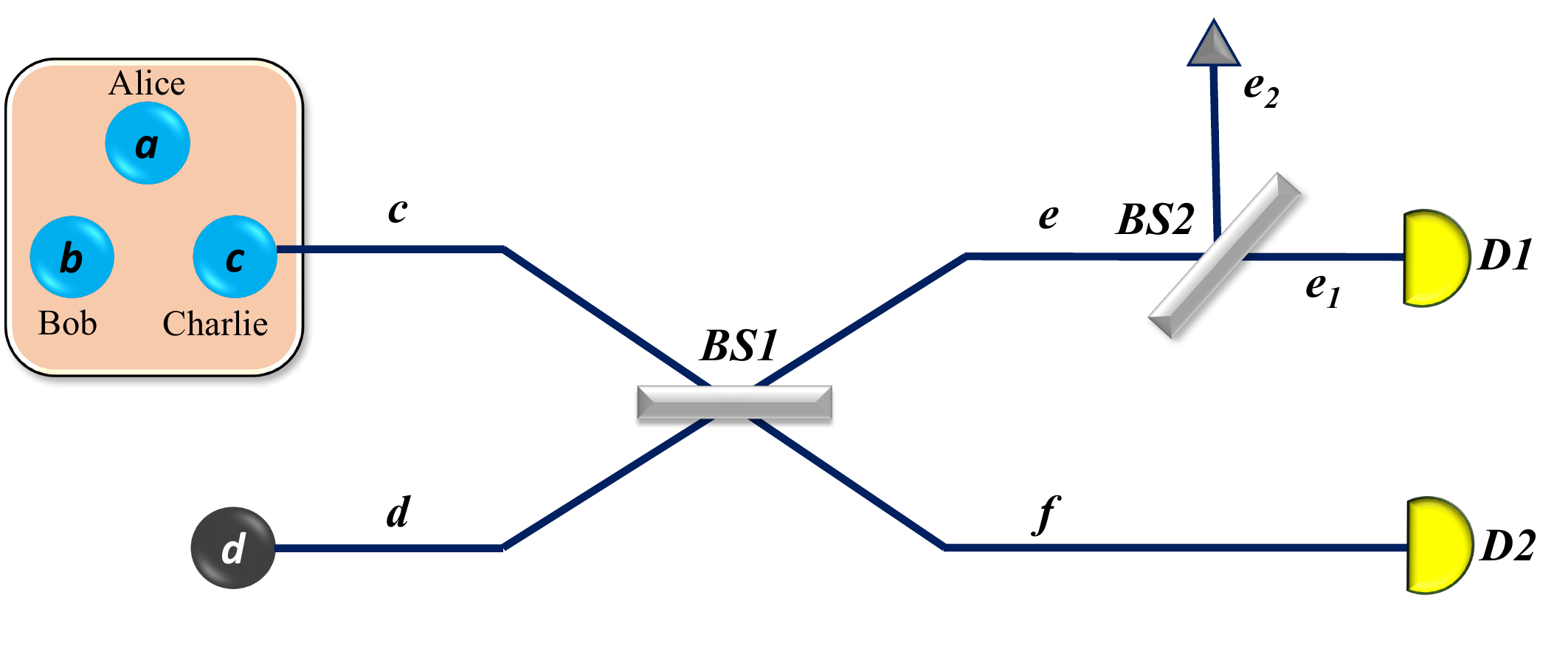}
\par\end{centering}
\caption{\label{fig:The-schematic-diagram}The schematic diagram of the proposed
ECP for 3-mode GHZ-type ECS. Initially, Alice, Bob and Charlie share
\textcolor{black}{the spacial modes $a$, $b$, and $c$, respectively}
of a partially entangled coherent state $|\Psi^{\prime}\rangle_{abc}$\textcolor{black}{{}
as }shown in Eq. \ref{eq:non}. Charlie also prepares \textcolor{black}{a
single-mode coherent state in spatial mode $d.$ Two beam splitters
$BS1$ and $BS2$ and two detectors $D1$ and $D2$ have been used
to perform the concentration protocol.}}

\end{figure}

\textcolor{black}{Let us suppose, while sharing/distribution (to perform
an application) of $a$, $b$ and $c$ modes, Eq.} \ref{eq:maximally-GHZ_ECS}\textcolor{black}{{}
gets transformed to a non-maximally ECS of the form  }

\textcolor{black}{
\begin{equation}
\begin{array}{ccc}
|\Psi^{\prime}\rangle_{abc} & = & N_{1}\left[\beta|\alpha\rangle_{a}|\alpha\rangle_{b}|\alpha\rangle_{c}+\gamma|-\alpha\rangle_{a}|-\alpha\rangle_{b}|-\alpha\rangle_{c}\right],\end{array}\label{eq:non}
\end{equation}
where $N_{1}=\left[\beta^{2}+\gamma^{2}+2\beta\gamma e^{-6|\alpha|^{2}}\right]^{-\frac{1}{2}}$is
the normalization coefficient. Here, $\beta$ and $\gamma$ are the
probability amplitudes considered to be real numbers.} In what follows,
we aim to achieve a maximally entangled 3-mode GHZ-type ECS (\ref{eq:maximally-GHZ_ECS})
from Eq. \ref{eq:non}. To do so, we perform ECP as shown in Fig.
\ref{fig:The-schematic-diagram}, where Alice, Bob and Charl\textcolor{black}{ie
share the non-maximally entangled ECS denoted in }Eq. \ref{eq:non}.\textcolor{black}{{}
In this ECP, first Charlie prepares a superposition of single-mode
coherent state in the spatial mode $d$ which can be expressed as }

\textcolor{black}{
\begin{equation}
\begin{array}{ccc}
|\varPhi\rangle_{d} & = & N_{2}\left[\gamma|\alpha\rangle_{d}+\beta|-\alpha\rangle_{d}\right],\end{array}\label{eq:single_coherent_state}
\end{equation}
where, $N_{2}=\left[\beta^{2}+\gamma^{2}+2\beta\gamma e^{-2|\alpha|^{2}}\right]^{-\frac{1}{2}}$
is the normalization coefficient. Now, the combined state can be written
as}

\textcolor{black}{
\begin{equation}
\begin{array}{lcl}
|\xi\rangle_{abcd} & = & |\varPsi^{\prime}\rangle_{abc}\otimes|\varPhi\rangle_{d}\\
|\xi\rangle_{abcd} & = & N_{1}\left[\beta|\alpha\rangle_{a}|\alpha\rangle_{b}|\alpha\rangle_{c}+\gamma|-\alpha\rangle_{a}|-\alpha\rangle_{b}|-\alpha\rangle_{c}\right]\otimes N_{2}\left[\gamma|\alpha\rangle_{d}+\beta|-\alpha\rangle_{d}\right]\\
|\xi\rangle_{abcd} & = & N_{1}N_{2}\left[\beta\gamma|\alpha\rangle_{a}|\alpha\rangle_{b}|\alpha\rangle_{c}|\alpha\rangle_{d}+\beta^{2}|\alpha\rangle_{a}|\alpha\rangle_{b}|\alpha\rangle_{c}|-\alpha\rangle_{d}\right.\\
 &  & \left.+\gamma^{2}|-\alpha\rangle_{a}|-\alpha\rangle_{b}|-\alpha\rangle_{c}|\alpha\rangle_{d}+\gamma\beta|-\alpha\rangle_{a}|-\alpha\rangle_{b}|-\alpha\rangle_{c}|-\alpha\rangle_{d}\right].
\end{array}\label{eq:combined}
\end{equation}
}

Subsequently,\textcolor{black}{{} they let the photons in spatial mode
$c$ and $d$ pass through the 50:50 beam splitter $(BS)$ denoted
as $BS1$. The $BS$ transforms two different coherent states $|\alpha\rangle$and
$|\beta\rangle$ to the states}

\begin{equation}
BS|\alpha\rangle|\beta\rangle\rightarrow\left|\frac{\alpha+\beta}{\sqrt{2}}\right\rangle \left|\frac{\alpha-\beta}{\sqrt{2}}\right\rangle .\label{eq:BS}
\end{equation}

There are four different possibilities exist when the photons in \textcolor{black}{spatial
mode $c$ and $d$ incident onto the 50:50 $BS$. It can be described
as}

\begin{equation}
\begin{array}{lcl}
BS|\alpha\rangle_{c}|\alpha\rangle_{d} & \rightarrow & |\sqrt{2}\alpha\rangle_{e}|0\rangle_{f},\\
BS|\alpha\rangle_{c}|-\alpha\rangle_{d} & \rightarrow & |0\rangle_{e}|\sqrt{2}\alpha\rangle_{f},\\
BS|-\alpha\rangle_{c}|\alpha\rangle_{d} & \rightarrow & |0\rangle_{e}|-\sqrt{2}\alpha\rangle_{f},\\
BS|-\alpha\rangle_{c}|-\alpha\rangle_{d} & \rightarrow & |-\sqrt{2}\alpha\rangle_{e}|0\rangle_{f}.
\end{array}\label{eq:possibilities}
\end{equation}

Following Eq. \ref{eq:possibilities}, the \textcolor{black}{photons
in spatial mode $c$ and $d$} (of Eq. \ref{eq:combined}) pass through
the $BS1$, then the Eq. \ref{eq:combined} can be expressed as

\textcolor{black}{
\begin{equation}
\begin{array}{ccc}
|\xi\rangle_{abef} & = & N_{1}N_{2}\left[\beta\gamma|\alpha\rangle_{a}|\alpha\rangle_{b}|\sqrt{2}\alpha\rangle_{e}|0\rangle_{f}+\beta^{2}|\alpha\rangle_{a}|\alpha\rangle_{b}|0\rangle_{e}|\sqrt{2}\alpha\rangle_{f}\right.\\
 & + & \left.\gamma^{2}|-\alpha\rangle_{a}|-\alpha\rangle_{b}|0\rangle_{e}|-\sqrt{2}\alpha\rangle_{f}+\gamma\beta|-\alpha\rangle_{a}|-\alpha\rangle_{b}|-\sqrt{2}\alpha\rangle_{e}|0\rangle_{f}\right].
\end{array}\label{eq:Post_BS}
\end{equation}
}

It is evident in the above equation that the terms $|\alpha\rangle_{a}|\alpha\rangle_{b}|\sqrt{2}\alpha\rangle_{e}|0\rangle_{f}$,
$|-\alpha\rangle_{a}|-\alpha\rangle_{b}|-\sqrt{2}\alpha\rangle_{e}|0\rangle_{f}$
and other terms $|\alpha\rangle_{a}|\alpha\rangle_{b}|0\rangle_{e}|\sqrt{2}\alpha\rangle_{f},\:|-\alpha\rangle_{a}|-\alpha\rangle_{b}|0\rangle_{e}|-\sqrt{2}\alpha\rangle_{f}$
do not have photons in the spatial mode $f$ and $e$, respectively.
Therefore, they adopt the post selection method to choose those two
terms where the spatial mode $f$ has no photon and the modified state
in Eq. \ref{eq:Post_BS} can be expressed (before normalization)

\textcolor{black}{
\begin{equation}
|\xi\rangle_{abe}=N_{1}N_{2}\left[\beta\gamma\left(|\alpha\rangle_{a}|\alpha\rangle_{b}|\sqrt{2}\alpha\rangle_{e}+|-\alpha\rangle_{a}|-\alpha\rangle_{b}|-\sqrt{2}\alpha\rangle_{e}\right)\right].\label{eq:post}
\end{equation}
}

\textcolor{black}{We can find that the state in Eq. \ref{eq:post}
has the same form with that in Eq. \ref{eq:maximally-GHZ_ECS}. However,
the amplitude of Eq. \ref{eq:post} in mode $e$ is $\sqrt{2}$ times
larger than the Eq. \ref{eq:maximally-GHZ_ECS} in spatial mode $c$,
which can be seen as an advantage for long-distance quantum communication.
Still, to obtain the desired maximally entangled 3-mode GHZ-type ECS
as shown in Eq. \ref{eq:maximally-GHZ_ECS}, the coherent state in
spatial mode $e$ passes through $BS2$ and the state evolves to}

\textcolor{black}{
\begin{equation}
|\xi\rangle_{abe_{1}e_{2}}=N_{1}N_{2}\left[\beta\gamma\left(|\alpha\rangle_{a}|\alpha\rangle_{b}|\alpha\rangle_{e_{1}}|\alpha\rangle_{e_{2}}+|-\alpha\rangle_{a}|-\alpha\rangle_{b}|-\alpha\rangle_{e_{1}}|-\alpha\rangle_{e_{2}}\right)\right].\label{eq:-2}
\end{equation}
}

To accomplish \textcolor{black}{our ECP,} Charlie performs the photon
number measurement of the coherent state in the mode $e_{1}$ without
distinguishing $|\pm\alpha\rangle$and the state becomes

\textcolor{black}{
\begin{equation}
|\xi\rangle_{abe_{2}}=N_{0}\left[|\alpha\rangle_{a}|\alpha\rangle_{b}|\alpha\rangle_{e_{2}}+|-\alpha\rangle_{a}|-\alpha\rangle_{b}|-\alpha\rangle_{e_{2}}\right].\label{eq:maximally_probability}
\end{equation}
Finally, they get the desired maximally entangled 3-mode GHZ-type
ECS which is of the form of Eq. \ref{eq:maximally-GHZ_ECS} with success
probability $P=2\left(N_{1}N_{2}\beta\gamma\right){}^{2}$.}

\section{\textcolor{black}{ECP }with two copies of partially entangled 3-mode
\textcolor{black}{GHZ-type ECS}s\label{sec:ECP-with-two}}

In this section, we have discussed how to obtain a maximally entangled
3-mode GHZ-type ECS in a conventional way, i.e., with the help of
two copies of partially entangled 3-mode GHZ-type ECSs.

\begin{figure}
\centering{}\includegraphics[scale=0.6]{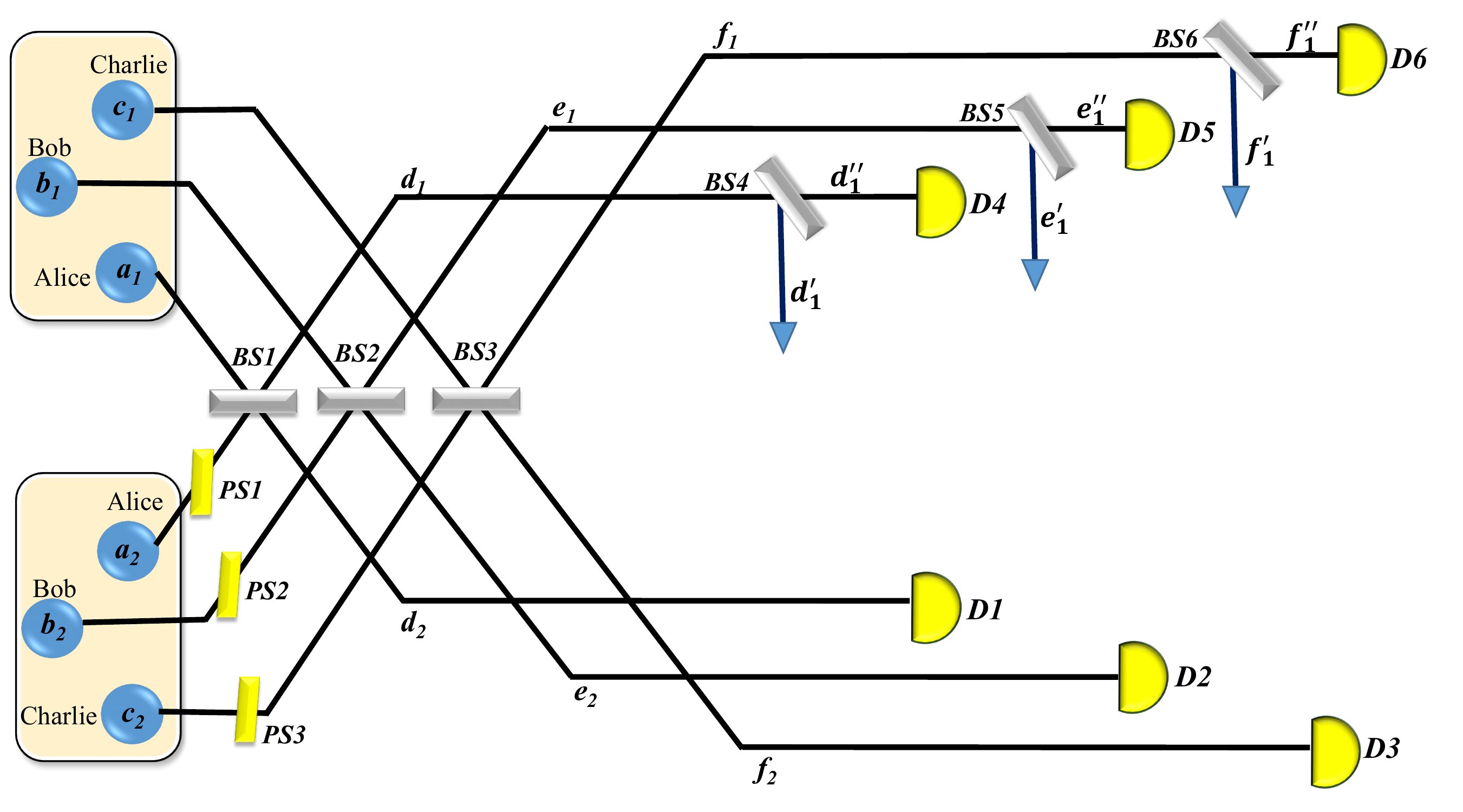}\caption{\label{fig:The-schematic-diagram-1}The schematic diagram of the proposed
ECP for 3-mode GHZ-type ECS. First, Alice, Bob and Charlie share two
copies of entangled coherent states \textcolor{black}{denoted as $a_{1},b_{1},c_{1}$
and $a_{2},b_{2},c_{2}$, respectively}. \textcolor{black}{Each of
them performs a phase shift $(PS1-PS3)$ operation on the second pair
($a_{2},b_{2},c_{2}$). Six beam splitters $BS1-BS6$ and six detectors
$D1-D6$ have been used to perform the concentration protocol.}}
\end{figure}

To begin with, Alice and Bob share two copies of partially entangled
3-mode GHZ-type ECSs according to Fig. \ref{fig:The-schematic-diagram-1},
which are of the form 

\begin{equation}
\begin{array}{lcl}
|\psi\rangle_{a_{1}b_{1}c_{1}} & = & N\left[\delta|\alpha\rangle_{a_{1}}|\alpha\rangle_{b_{1}}|\alpha\rangle_{c_{1}}+\eta|-\alpha\rangle_{a_{1}}|-\alpha\rangle_{b_{1}}|-\alpha\rangle_{c_{1}}\right],\end{array}\label{eq:one_copy}
\end{equation}

and

\begin{equation}
\begin{array}{lcl}
|\psi\rangle_{a_{2}b_{2}c_{2}} & = & N\left[\delta|\alpha\rangle_{a_{2}}|\alpha\rangle_{b_{2}}|\alpha\rangle_{c_{2}}+\eta|-\alpha\rangle_{a_{2}}|-\alpha\rangle_{b_{2}}|-\alpha\rangle_{c_{2}}\right].\end{array}\label{eq:Second_copy}
\end{equation}

First copy is in spatial mode $a_{1},b_{1},c_{1}$ while the second
copy is in spatial mode $a_{2},b_{2},c_{2}.$ In Eqs. \ref{eq:one_copy}
and \ref{eq:Second_copy}, $N=\left[\delta^{2}+\eta^{2}+2\delta\eta e^{-6|\alpha|^{2}}\right]^{-\frac{1}{2}}$
represents the normalization coefficient, whereas we have considered
$\delta$ and $\eta$ are real numbers. Further, to initiate the concentration
process, first Alice, Bob and Charlie perform a phase shift $(PS)$
operation on their respective modes $a_{2},b_{2},$ and $c_{2}$ corresponding
to the second copy in Eq. \ref{eq:Second_copy}. Consequently, the
state in spatial mode $a_{2},b_{2},c_{2}$ becomes

\begin{equation}
\begin{array}{lcl}
|\psi^{\prime}\rangle_{a_{2}b_{2}c_{2}} & = & N\left[\eta|\alpha\rangle_{a_{2}}|\alpha\rangle_{b_{2}}|\alpha\rangle_{c_{2}}+\delta|-\alpha\rangle_{a_{2}}|-\alpha\rangle_{b_{2}}|-\alpha\rangle_{c_{2}}\right].\end{array}\label{eq:after_ps}
\end{equation}

The combined state $|\psi\rangle_{a_{1}b_{1}c_{1}}$ and $|\psi^{\prime}\rangle_{a_{2}b_{2}c_{2}}$
can be expressed as 

\begin{equation}
\begin{array}{cl}
 & |\chi\rangle_{a_{1}b_{1}c_{1}a_{2}b_{2}c_{2}}\\
= & |\psi\rangle_{a_{1}b_{1}c_{1}}\otimes|\psi^{\prime}\rangle_{a_{2}b_{2}c_{2}}\\
= & N\left[\delta|\alpha\rangle_{a_{1}}|\alpha\rangle_{b_{1}}|\alpha\rangle_{c_{1}}+\eta|-\alpha\rangle_{a_{1}}|-\alpha\rangle_{b_{1}}|-\alpha\rangle_{c_{1}}\right]\otimes N\left[\eta|\alpha\rangle_{a_{2}}|\alpha\rangle_{b_{2}}|\alpha\rangle_{c_{2}}+\delta|-\alpha\rangle_{a_{2}}|-\alpha\rangle_{b_{2}}|-\alpha\rangle_{c_{2}}\right]\\
= & N^{2}\left[\delta\eta|\alpha\rangle_{a_{1}}|\alpha\rangle_{b_{1}}|\alpha\rangle_{c_{1}}|\alpha\rangle_{a_{2}}|\alpha\rangle_{b_{2}}|\alpha\rangle_{c_{2}}+\delta^{2}|\alpha\rangle_{a_{1}}|\alpha\rangle_{b_{1}}|\alpha\rangle_{c_{1}}|-\alpha\rangle_{a_{2}}|-\alpha\rangle_{b_{2}}|-\alpha\rangle_{c_{2}}\right.\\
 & \left.+\eta^{2}|-\alpha\rangle_{a_{1}}|-\alpha\rangle_{b_{1}}|-\alpha\rangle_{c_{1}}|\alpha\rangle_{a_{2}}|\alpha\rangle_{b_{2}}|\alpha\rangle_{c_{2}}+\delta\eta|-\alpha\rangle_{a_{1}}|-\alpha\rangle_{b_{1}}|-\alpha\rangle_{c_{1}}|-\alpha\rangle_{a_{2}}|-\alpha\rangle_{b_{2}}|-\alpha\rangle_{c_{2}}\right].
\end{array}\label{eq:combined-1}
\end{equation}

Now, the spatial modes $a_{1}a_{2},$ $b_{1}b_{2}$ and $c_{1}c_{2}$
will pass through the $BS1$, $BS2$ and $BS3$, respectively. Similar
to Eq. \ref{eq:Post_BS}, by following the Eq. \ref{eq:possibilities}
the state becomes

\begin{equation}
\begin{array}{lcl}
|\chi\rangle_{d_{1}d_{2}e_{1}e_{2}f_{1}f_{2}} & = & N^{2}\left[\delta\eta|\sqrt{2}\alpha\rangle_{d_{1}}|0\rangle_{d_{2}}|\sqrt{2}\alpha\rangle_{e_{1}}|0\rangle_{e_{2}}|\sqrt{2}\alpha\rangle_{f_{1}}|0\rangle_{f_{2}}\right.\\
 &  & +\delta^{2}|0\rangle_{d_{1}}|\sqrt{2}\alpha\rangle_{d_{2}}|0\rangle_{e_{1}}|\sqrt{2}\alpha\rangle_{e_{2}}|0\rangle_{f_{1}}|\sqrt{2}\alpha\rangle_{f_{2}}\\
 &  & +\eta^{2}|0\rangle_{d_{1}}|-\sqrt{2}\alpha\rangle_{d_{2}}|0\rangle_{e_{1}}|-\sqrt{2}\alpha\rangle_{e_{2}}|0\rangle_{f_{1}}|-\sqrt{2}\alpha\rangle_{f_{2}}\\
 &  & \left.+\delta\eta|-\sqrt{2}\alpha\rangle_{d_{1}}|0\rangle_{d_{2}}|-\sqrt{2}\alpha\rangle_{e_{1}}|0\rangle_{e_{2}}|-\sqrt{2}\alpha\rangle_{f_{1}}|0\rangle_{f_{2}}\right].
\end{array}\label{eq:after_BS}
\end{equation}

With the post selection method, they choose the items $|\sqrt{2}\alpha\rangle_{d_{1}}|0\rangle_{d_{2}}|\sqrt{2}\alpha\rangle_{e_{1}}|0\rangle_{e_{2}}|\sqrt{2}\alpha\rangle_{f_{1}}|0\rangle_{f_{2}}$
and $|-\sqrt{2}\alpha\rangle_{d_{1}}|0\rangle_{d_{2}}|-\sqrt{2}\alpha\rangle_{e_{1}}|0\rangle_{e_{2}}|-\sqrt{2}\alpha\rangle_{f_{1}}|0\rangle_{f_{2}}$
in which the spatial mode $d_{2},$$e_{2}$ and $f_{2}$ have no photon.
The state evolves to

\begin{equation}
\begin{array}{lcl}
|\chi\rangle_{d_{1}e_{1}f_{1}} & = & N^{2}\left[\delta\eta\left(|\sqrt{2}\alpha\rangle_{d_{1}}|\sqrt{2}\alpha\rangle_{e_{1}}|\sqrt{2}\alpha\rangle_{f_{1}}\right.\right.\\
 &  & \left.\left.+|-\sqrt{2}\alpha\rangle_{d_{1}}|-\sqrt{2}\alpha\rangle_{e_{1}}|-\sqrt{2}\alpha\rangle_{f_{1}}\right)\right].
\end{array}\label{eq:-1}
\end{equation}

The above equation is same as Eq. \ref{eq:maximally-GHZ_ECS} with
the only difference that the amplitude in modes $d_{1},e_{1}$ and
$f_{1}$ is amplified (an advantage in long-distance quantum communication)
by $\sqrt{2}$ in comparison to modes $a,b$ and $c$ of Eq. \ref{eq:maximally-GHZ_ECS}.
Therefore, as shown in Fig. \ref{fig:The-schematic-diagram-1}, they
use the beamsplitters $BS4$, $BS5$ and $BS6$ to pass the spatial
mode $d_{1},$$e_{1}$ and $f_{1}$ of $|\chi\rangle_{d_{1}e_{1}f_{1}}$,
and the state evolved as 

\begin{equation}
\begin{array}{lcl}
|\chi\rangle_{d_{1}^{\prime}d_{1}^{\prime\prime}e_{1}^{\prime}e_{1}^{\prime\prime}f_{1}^{\prime}f_{1}^{\prime\prime}} & = & N^{2}\left[\delta\eta\left(|\alpha\rangle_{d_{1}^{\prime}}|\alpha\rangle_{d_{1}^{\prime\prime}}|\alpha\rangle_{e_{1}^{\prime}}|\alpha\rangle_{e_{1}^{\prime\prime}}|\alpha\rangle_{f_{1}^{\prime}}|\alpha\rangle_{f_{1}^{\prime\prime}}\right.\right.\\
 &  & \left.\left.+|-\alpha\rangle_{d_{1}^{\prime}}|-\alpha\rangle_{d_{1}^{\prime\prime}}|-\alpha\rangle_{e_{1}^{\prime}}|-\alpha\rangle_{e_{1}^{\prime\prime}}|-\alpha\rangle_{f_{1}^{\prime}}|-\alpha\rangle_{f_{1}^{\prime\prime}}\right)\right].
\end{array}\label{eq:after_BS-1}
\end{equation}

Subsequently, in order to complete \textcolor{black}{our ECP,} Alice,
Bob and Charlie perform the photon number measurement of their respective
coherent state in the mode $d_{1}^{\prime\prime},e_{1}^{\prime\prime}$
and $f_{1}^{\prime\prime}$ without distinguishing $|\pm\alpha\rangle$and
the state can be written as 

\begin{equation}
\begin{array}{lcl}
|\chi\rangle_{d_{1}^{\prime}e_{1}^{\prime}f_{1}^{\prime}} & = & N_{0}\left[|\alpha\rangle_{d_{1}^{\prime}}|\alpha\rangle_{e_{1}^{\prime}}|\alpha\rangle_{f_{1}^{\prime}}+|-\alpha\rangle_{d_{1}^{\prime}}|-\alpha\rangle_{e_{1}^{\prime}}|-\alpha\rangle_{f_{1}^{\prime}}\right].\end{array}\label{eq:Final_MES_GHZ_ECS}
\end{equation}

The above equation is our desired state as shown in Eq. \ref{eq:maximally-GHZ_ECS}.
Finally, they obtain the concentrated maximally entangled 3-mode GHZ-type
ECS with the total success probability $P=2\left(N^{2}\delta\eta\right){}^{2}.$

\section{Success probability\label{sec:Success-probability}}

\begin{figure}[H]
\begin{centering}
\includegraphics[scale=0.5]{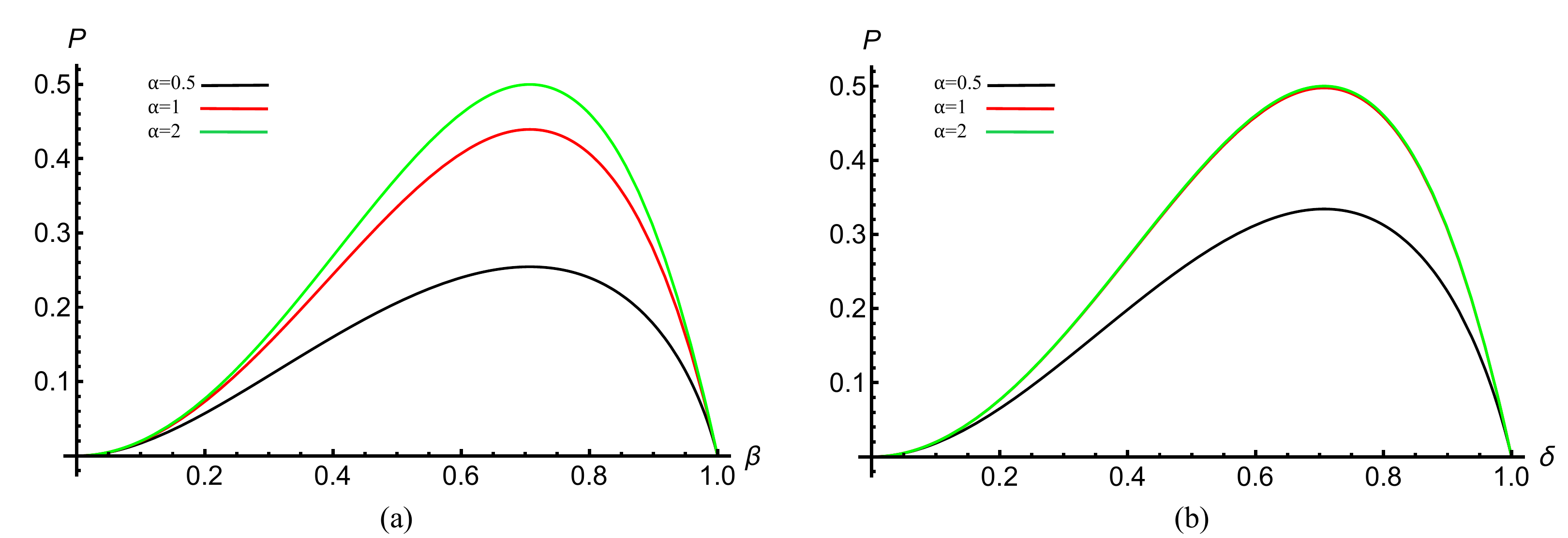}
\par\end{centering}
\caption{\label{fig:(Color-online)-Variation}(Color online) Variation of success
probability $(P)$ to obtain a maximally entangled 3-mode GHZ-type
ECS. (a) The variation of $P$ with the coefficient $\beta$, (b)
The variation of $P$ with the coefficient $\delta$, assuming $\alpha=0.5,$
$1,$ and $2$. }
\end{figure}

In the last two sections, we have discussed the two ECPs for partially
entangled 3-mode GHZ-type ECS. According to our first and second ECP
performed in Sec. \ref{sec:ECP-for-GHZ-type} and Sec. \ref{sec:ECP-with-two},
we can calculate the success probabilities \textcolor{black}{$P=2\left(N_{1}N_{2}\beta\gamma\right){}^{2}$
and $P=2\left(N^{2}\delta\eta\right){}^{2}$, respectively. Now, we
have plotted both of these $P$ in Fig.} \ref{fig:(Color-online)-Variation}
(a) and (b) and discussed the variation of $P$ with state parameters.
Specifically, in Fig. \ref{fig:(Color-online)-Variation} (a), variation
of $P$ with $\beta$ has been shown for $\alpha=0.5$, $1$, and
$2$, which manifests that the success probability $P$ increases
with the increase in $\alpha$ and reaches at the peak (maximum possible
value) at a constant value of $\beta=0.7$ for all values of $\alpha$.
In Fig. \ref{fig:(Color-online)-Variation} (b), variation of $P$
with $\delta$ has been shown for $\alpha=0.5$, $1$, and $2$, which
reveals exactly the similar behaviors as in Fig. \ref{fig:(Color-online)-Variation}
(a), though, the peak value is more i.e., $P=3$ for $\alpha=0.5$
and $P=4.9$ for $\alpha=1$ than the case in Fig. \ref{fig:(Color-online)-Variation}
(a). However, for $\alpha=2$, both Fig. \ref{fig:(Color-online)-Variation}
(a) and (b) are equivalent. Further, our first ECP is more simple
and only Charlie needs to perform the operation. On the other hand,
our second ECP has an advantage as it is more efficient than our first
ECP for $\alpha<2$. It is to be noted that Ref. \cite{ECP_Bell_type_ECS}
would also obtain the similar nature as we have shown in Fig. \ref{fig:(Color-online)-Variation}.

Furthermore, as there are two non-equivalent classes for 3-qubit entangled
states i.e., GHZ class and W class \cite{GHZ-W-Class}, so it would
be interesting to compare the success probabilities $P$ of our first
ECP (Sec. \ref{sec:ECP-for-GHZ-type}) for 3-mode GHZ-type ECS as
shown in Fig. \ref{fig:(Color-online)-Variation} (a) with 3-mode
W-type ECS proposed in Sec. 2 and its corresponding Fig. 3 of \cite{W-ECS}.
It is worth mentioning that our ECP for 3-mode GHZ-type ECS exhibits
maximal $P$ at $\beta=0.7$ for $\alpha=0.5,1$ and $2$. However,
for 3-mode W-type ECS \cite{W-ECS} $P$ reaches at the maximal value
at $\gamma=0.5$ for $\alpha=0.5$ and at $\gamma=0.65$ for $\alpha=1,2$. 

Although for $\alpha=0.5$, our ECP for 3-mode GHZ-type ECS has slightly
less maximal value $(P=2.5)$ than for 3-mode W-type ECS \cite{W-ECS}
maximal value $(P=3)$, however, our ECP has peak for larger value
of $\beta=0.7$ than for \cite{W-ECS} which has peak for $\gamma=0.5$.
Similarly for $\alpha=1$, our ECP for 3-mode GHZ-type ECS has slightly
less maximal value $(P=4.5)$ than for 3-mode W-type ECS \cite{W-ECS}
maximal value $(P=4.8)$, however, our ECP has peak for larger value
of $\beta=0.7$ than for \cite{W-ECS} which has peak for $\gamma=0.64$.
Subsequently for $\alpha=2$, our ECP for 3-mode GHZ-type ECS has
the same maximal value $(P=5)$ as for 3-mode W-type ECS \cite{W-ECS}
maximal value $(P=5)$, however, our ECP has peak for larger value
of $\beta=0.7$ than for \cite{W-ECS} which has peak for $\gamma=0.63$.
Therefore, our ECP for 3-mode GHZ-type ECS as shown in Fig. \ref{fig:(Color-online)-Variation}
(a) is more efficient (has maximal $P$ values) for larger values
of state parameter than for 3-mode W-type ECS shown in Fig. 3 of \cite{W-ECS}. 

\section{Conclusion\label{sec:Conclusion}}

Many quantum communication and technology applications which also
have been mentioned in the introduction section, require distributed
(shared) entanglement (maximal) that degrades due to decoherence while
processing, manipulating and storing the quantum systems. Hence the
degraded entanglement needs to be enhanced, however, it cannot be
increased by local operation and classical communication. Consequently,
Alice and Bob usually share a mixed entangled state which results
in the fidelity less than unity. The only solution to this is to perform
an ECP/EPP prior to start any quantum communication or computation
task. 

Here, we aim to propose the ECPs for 3-mode GHZ-type ECS for its use
in the various applications proposed in the past \cite{QKD_GHZ_ECS}
and in recent years \cite{H_Prakash_2019_CED,H_Prakash_2019_CQT}.
Basically, the development of ECSs based applications \cite{ECS_Tele_2,ECS_Tele_3,ECS_Tele_4}
are dependent on the proposal of the corresponding ECPs so that their
physical realization can take place with unit success probability.
On account of the importance of such CV applications, we have proposed
two ECPs. In Sec. \ref{sec:ECP-for-GHZ-type}, the first ECP uses
a superposition of single-mode coherent state and in Sec. \ref{sec:ECP-with-two}
the second ECP uses the conventional way requiring the two copies
of partially entangled 3-mode GHZ-type ECSs to obtain the concentrated
maximally entangled 3-mode GHZ-type ECS. Further, we have discussed
their success probabilities $P$ in detail and later we have compared
the success probabilities of our first ECP for 3-mode GHZ-type ECS
with an ECP of 3-mode W-type ECS and found that our ECP is more efficient
(maximal success probabilities) for larger value $(\beta=0.7)$ of
state parameter. Finally, we have concluded our work with the expectation
that our ECPs can be implemented experimentally in the near future
for long-distance quantum communication or computation tasks. 

\textbf{Acknowledgments: }MS thanks to the Department of Science and
Technology (DST), India, for support provided through the DST project
No. S/DST/VNA/20190004. Authors also thank Kishore Thapliyal for his
interest in the work and helpful suggestions.

\end{document}